\documentclass[11pt, a4paper]{article}

\usepackage{amsmath,amssymb, amsthm}
\usepackage{inputenc}
\usepackage{mathrsfs}
\usepackage{fullpage}
\usepackage[affil-it]{authblk}

\usepackage{graphicx}
\graphicspath{ {images/} }

\usepackage{wrapfig}

\usepackage{dsfont}
\usepackage{tikz}

\usetikzlibrary{arrows,calc}

\usepackage{verbatim}

\usepackage{hyperref}

\usepackage{xcolor}
\hypersetup{
  colorlinks   = true, 
  urlcolor     = {blue!90!black}, 
  linkcolor    = {blue!90!black}, 
  citecolor   = {red!90!black} 
}

\usepackage{geometry}   
\usepackage{amsmath}
\usepackage{xcolor}
\usepackage{graphicx}
\usepackage{amssymb}
\usepackage{epstopdf}
\usepackage{enumerate}
\usepackage{bbm}
\usepackage{braket}

\newcommand{\R}{\mathbb{R}}
\newcommand{\N}{\mathbb{N}}
\newcommand{\Z}{\mathbb{Z}}

\newcommand{\Y}{\mathcal{Y}}

\newcommand{\EE}[1]{\mathbb{E}\left(#1\right)}
\newcommand{\E}{\mathbb{E}}
\renewcommand{\P}{\mathbb{P}}
\newcommand{\PP}[1]{\mathbb{P}\left\{#1\right\}}

\theoremstyle{definition}
\newtheorem{defi}{Definition}[section]

\theoremstyle{plain}
\newtheorem{thm}[defi]{Theorem}

\newtheorem{lem}[defi]{Lemma}
\newtheorem{cor}[defi]{Corollary}

\theoremstyle{definition}

\author[1]{Viktor Bezborodov \thanks{Email: \texttt{vbezborodov@math.uni-bielefeld.de}}} 
\author[1]{
Luca Di Persio \thanks{Email: \texttt{dipersioluca@gmail.com}}}

\affil[1]{
\emph{The University of Verona}}

\title{The quenched central limit theorem for a  model of random walk in random environment}

\begin{document}

\maketitle

\begin{abstract}

A short proof of the quenched central limit theorem for 
the random walk in random environment introduced 
by 
 Boldrighini,  Minlos, and Pellegrinotti \cite{BMP94}
 is given.
 
\end{abstract}

\section{Introduction}

In this article we give a short proof of the quenched 
central limit theorem for a model 
of the random walk in random environment.
At each site
the transition probability kernel is affected
by the current state of the environment 
at this site.
The model was introduced 
 by Boldrighini,  Minlos, and Pellegrinotti,
 see in particular \cite{BMP94, BMP97, BMP07}
 and more recent papers
 Boldrighini et al. \cite{BMPZ15} and
 Di Persio \cite{DP10}.
 The model is described in Section \ref{sec2}.
 Boldrighini et al. \cite{BMPZ15} contains 
 a nice overview of the literature on the subject.
 For a survey on the recent progress on this and similar 
 models see Zeitouni \cite{Zei06}
 or Biskup \cite{surv}. 
 A related model is considered by
 Barraquand
and Corwin \cite{BC16} and
Thiery and Le Doussal \cite{TL17}.

 The proof makes use of 
 the multidimensional martingale CLT
 by  K{\"u}chler and S{\o}rensen
 \cite{KS99}. The paper is organized as follows.
 In Section \ref{sec2} we describe the model 
 and give the statement. 
 In Section \ref{secproofs} we give the proofs
 and some further comments.

\section{Model, conditions and results}\label{sec2}

Consider a particle moving in a $n$-dimensional infinite lattice and denote by $X_t$ is position at time $t$.
On the lattice, a dynamical random environment is considered. It is described by the random field
\[
\xi=\set{\xi_t(x):x\in\mathbb{Z}^n, t\in\mathbb{Z}^+}
\]
Note that the time  is discrete.
We assume that $\xi$ is the result of 
independent copies of the same random variable
taking values in some finite space $\mathbb{S}$.
The space of configurations is given by 
$\tilde \Omega=\mathbb{S}^{\mathbb{Z}^n\times \mathbb{Z}^+}$.
The values of the field for each $(t,x)\in\mathbb{Z}^{n+1}$, i.e. $\set{\xi_t(x)}_{(x,t)\in\mathbb{Z}^{n+1}}$, are i.i.d random variables, distributed according to a given probability measure denoted by $\pi$.

The one step transition probability from position $x$ at same time $t$ to position $y$ at the subsequent time step is given by
\[
	\PP {X_{t+1}=y\lvert X_t=x,\xi}=P_0(y-x)+ c(y-x,\xi_t(x))
\]
where $P_0$ is the transition probability of a free random walk and $c$ is the function which provides the influence of the environment on the particle's dynamic.

In order for the probability $P$ to be well-defined, the following conditions must be fulfilled:
\begin{itemize}
\item
$0\le P_0(u)+ c (u,s)\le 1\quad\forall s\in\mathbb{S}\quad\forall u\in\mathbb{Z}^n$;
\item
$\sum_{u\in\mathbb{Z}^n} c(u,s)=0\quad\forall s\in\mathbb{S}$.
\end{itemize}

Moreover we assume that the random environment has the following property:
\begin{equation}
  \sum_{s\in\mathbb{S}} c(u,s)\pi(s)=0\text{ for any }u\in\mathbb{Z}^n ,
\end{equation}
 which
 means that $P_0$ is the mean transition probability.

 Additionally, let $P_0$ and $c$ be of bounded range.
 We denote by $\P _ \xi$ be 
 the conditional probability with respect to the environment.
 
 Let us define the `average' transition probability 

\begin{equation}
 \bar P(u) = P_0(u) + 
 \sum\limits _{s\in\mathbb{S}} \pi (s)c(u,s).
\end{equation}
 
 We further assume that for some $b^c \in \R$,
 \begin{equation} \label{encroach}
  \sum_{u\in\mathbb{Z}^n} u c(u,s)=b^c \quad\forall s\in\mathbb{S}.
 \end{equation}

Let $Y=\set{Y_t}_{t\in\mathbb{Z}^+}$ 
be the stochastic processes defined by $Y_t=X_t-tb$,
where $b = b^0 + b^c$ and $b^0 = \sum_{u\in\mathbb{Z}^n} u P_0 (u)$.
 Note that 
 \[
  \sum_{u\in\mathbb{Z}^n} (u - b)
  \left[ P_0(u)+ c(u,s) \right] = 0 \quad\forall s\in\mathbb{S}.
 \]

 \begin{thm}\label{bodacious}
 For almost every  realisation $\xi$
 of the random environment
 we have
 \begin{equation}
  \frac{1}{t}Y_t \Rightarrow \eta ^2 U,
 \end{equation}
  $\P _ \xi $ -a.s.,
   where
   $U$ is a standard normal vector 
   and $\eta ^2$ is the positive semidefinite matrix with entries
\begin{equation}\label{insidious}
  (\eta ^2)_{ij} = 
   \sum\limits _{u \in \Z ^n}
  ( u_i -b_i) ( u_j - b_j)  \bar P (u).
\end{equation}
 \end{thm}

 \section{Proofs}\label{secproofs}

\begin{lem}
For every $\xi \in \tilde \Omega$,
  the process $Y$ is a martingale
under $\P_\xi$.
\end{lem}
\textbf{Proof}.
This is a direct consequence of
the definition of $Y$ and
\eqref{encroach}.

 Define $H _t = \EE{Y_t Y _t '}$, where $Y _t '$ the 
 transposed matrix, and the matrix 
 $[Y]_t = ([Y^i, Y^j]_t)_{1\leq i,j \leq n}$, 
 and $H _t ^\xi = \E _\xi (Y_t Y _t ')$.
 Let also $K_t = \frac{1}{\sqrt{t}} I_n$,
 where $I_n$ is the $n\times n$ identity matrix.
 
 \begin{lem}
 We have
  \begin{equation}\label{lore}
  \begin{gathered}
   \E _ \xi \left[ (Y^i_{r+1} - Y ^i _r)(Y^j_{r+1} - Y ^j _r) 
   \right] 
   \\ =
   \sum\limits _{y \in \Y} \PP {Y  _r = y \mid  \xi }
   \sum\limits _{u \in \Z ^n}
  ( u_i -b_i) ( u_j - b_j)  \left[ P_0(u) + c(u,\xi_r(y)) \right].
  \end{gathered}
  \end{equation}
The above sum by $y$ is taken over the countable set 
$$
 \Y := \{ z_1 + z_2 b \mid z_1, z_2 \in \Z \}.
$$
(Note that $\PP{ Y_t \in \Y \text{ for all } t \in \N} = 1$).
  
 \end{lem}

 \textbf{Proof}.
 By definition of $Y$ and $\P_ \xi$,
 \[
  \E _ \xi \left[ (Y^i_{r+1} - Y ^i _r)(Y^j_{r+1} - Y ^j _r) 
   \right] 
 \]
 \[ =
  \E\left[ (Y^i_{r+1} - Y ^i _r)(Y^j_{r+1} - Y ^j _r) 
   \middle| \xi \right] = 
   \E\left[ \E\left\{ 
  (Y^i_{r+1} - Y ^i _r)(Y^j_{r+1} - Y ^j _r) \middle|
  \xi, Y_r\right\}
   \middle| \xi \right]
 \]
\[
 = \E \left[ \sum\limits _{u} 
 ( u_i -b_i - Y^i _r) ( u_j - b_j- Y^j _r)
 [P_0(u - Y_r) + c(u - Y_r, \xi(Y_r)] \middle| \xi \right]
\]
\[
  = \E\left[ \sum\limits _{u} 
 ( u_i -b_i ) ( u_j - b_j)
 [P_0(u ) + c(u, \xi(Y_r)] \middle| \xi \right]
\]
\[
 = \sum\limits _{y \in \mathcal{Y}} \P \{Y  _r = y \mid  \xi \}
   \sum\limits _{u \in \Z ^n}
  ( u_i -b_i) ( u_j - b_j)  \left[ P_0(u) + c(u,\xi_r(y)) \right].
\]
 \qed

 \begin{lem} \label{assuage}
 We have
  \begin{equation}
   \frac{1}{t} [M]_t \to \eta ^2,
  \end{equation}
 $P_\xi$-a.s. for a.a. $\xi$.

 \end{lem}
 
 \textbf{Proof}.
 Note that  
 for  $1 \leq i, j \leq n$,
 
 \[
  ([Y]_t)_{ij} =
  \sum\limits _{0\leq r < t}
  \Delta _{r, ij},
 \]
where
\[
 \Delta _{r, ij} =  \big[Y ^i _{r+1} -
   Y ^i _r \big] \big[
  Y ^j _{r+1} -
   Y ^j _r  \big].
\]

  Under $\P _\xi$ a.s. on $\{ Y_t = y \}$ the distribution 
  of $Y  _{t+1} -
   Y  _t $ is $P_0(u) + c(u, \xi_t(y))$. 
   Since under $\P _\xi$ the random variables $  Y  _{t+1} -
   Y  _t  $ are independent of each other for 
   different $t$,
  the statement of the lemma follows
  from the law of large numbers.
  \qed
  
  \begin{cor} Lemma
  \ref{assuage} also holds $P$-a.s.
  \end{cor}

 \begin{lem}\label{rapport}
 \begin{itemize}
 \item[]$(i)$ We have
  \begin{equation}
  (H_{r+1})_{ij} -  (H_{r})_{ij} = 
  \sum\limits _{s\in\mathbb{S}} \pi (s)
   \sum\limits _{u \in \Z ^n}
  ( u_i -b_i) ( u_j - b_j)  \left[ P_0(u) + c(u,s) \right].
\end{equation}
\item[]$(ii)$ We also have
 \begin{equation}
 \begin{gathered}
  (H ^ {\xi} _{r+1})_{ij} -  (H^ {\xi} _{r})_{ij}  \\ = 
  \sum\limits _{y \in \Y} \P _{\xi}\{ Y _r = y\}
   \sum\limits _{u \in \Z ^n}
  ( u_i -b_i) ( u_j - b_j)  \left[ P_0(u) + c(u,\xi_r(y)) \right].
   \end{gathered}
\end{equation}
\end{itemize}
 \end{lem}

 \textbf{Proof} $(i)$ We start by noting that for $i,j \in \{1,...,n\}$,
 
 \begin{equation}\label{dibber}
    \E \left( (Y^i_{t+1} -Y^i_{t}) Y^j_{t} \right) = 0.
 \end{equation}
Indeed,
\[
 \E \left( (Y^i_{t+1} -Y^i_{t}) Y^j_{t} \right) = 
 \E \E \left[ (Y^i_{t+1} - Y^i_{t}) Y^j_{t} \middle| Y_t \right]
\]
\[
 = \sum\limits _{y \in \Y} \P\{Y_t = y \} \sum\limits _{u \in \Z ^n}
 (y_i + u _i - b _i - y _i)y _j \bar P(u) = 
 \sum\limits _{y \in \Y} \P\{Y_t = y \} y _j \sum\limits _{u \in \Z ^n}
 (u _i - b _i ) \bar P (u) = 0.
\]

By \eqref{dibber},
\[
 (H_{r+1})_{ij} -  (H_{r})_{ij} = 
 \E\left( Y^i_{t+1}Y^j _{t+1} - Y^i_{t}Y^j _{t} \right)
\]
\[
 = \E\left( (Y^i_{t+1}- Y^i_{t}  )
 (Y^j _{t+1} - Y^j _{t}) \right) + 
 \E\left( (Y^i_{t+1}- Y^i_{t}  )
  Y^j _{t} \right)+
 \E\left(  Y^i_{t}  
 (Y^j _{t+1} - Y^j _{t}) \right)
\]
\[
  = \E\left( (Y^i_{t+1}- Y^i_{t}  )
 (Y^j _{t+1} - Y^j _{t}) \right).
\]
Conditioning on $Y_t$, we get 
 \[
  (H_{r+1})_{ij} -  (H_{r+1})_{ij} = 
  \sum\limits _y P \{Y_t = y \}
  \sum\limits _{u} (u_i - b_i)(u_j - b_j) 
  \left[ P_0(u) + c(u,\xi _r (y)) \right]
  \]
  \[
  =
  \sum\limits _{u} (u_i - b_i)(u_j - b_j) \bar P (u).
 \]

 $(ii)$ \eqref{dibber} holds for $\E_\xi$ too, since
 
 \[
   \E \E \left[ (Y^i_{t+1} - Y^i_{t}) Y^j_{t} \middle| Y_t, \xi \right]
 \]
\[
 = \sum\limits _{y \in \Y} \P\{Y_t = y \} \sum\limits _{u \in \Z ^n}
 (y_i + u _i - b _i - y _i)y _j [ P _0 (u) + c(u, \xi_t (y))]
\]
\[
 =\sum\limits _{y \in \Y} \P\{Y_t = y \} y _j \sum\limits _{u \in \Z ^n}
 ( u _i - b _i ) P _0 (u) 
 +\sum\limits _{y \in \Y} \P\{Y_t = y \} y _j \sum\limits _{u \in \Z ^n}
  u _i    c(u, \xi_t (y)) 
  \]
  \[
  - b _i \sum\limits _{y \in \Y} \P\{Y_t = y \} y _j \sum\limits _{u \in \Z ^n}
    c(u, \xi_t (y)) 
\]
\[
 = \sum\limits _{y \in \Y} \P\{Y_t = y \} ( - y _j b^c)
 + \sum\limits _{y \in \Y} \P\{Y_t = y \} y _j b^c
 -0 = 0.
\]

The proof continues as in $(i)$.
 \qed

 \begin{lem}\label{pull in horns}
  \begin{equation}\label{detriment}
   \frac{1}{t} H_t \to \eta ^2, \ \ \
   \frac{1}{t} H ^ {\xi}_t \to \eta ^2,
  \end{equation}
where $\eta ^2$ is as in \eqref{insidious},
$\P _\xi$-a.s 
for $\Pi$-a.a. $\xi $.
 \end{lem}
 
 \textbf{Proof}.
 Let us only prove the second convergence in 
 \eqref{detriment}. By Lemma \ref{rapport},
 
 \[
 (H ^ {\xi}_t)_{ij} = \sum\limits _{r=0}^{t-1}
  \sum\limits _{y \in \Y} \P _{\xi}\{ Y _r = y\}
   \sum\limits _{u \in \Z ^n}
  ( u_i -b_i) ( u_j - b_j)  \left[ P_0(u) + c(u,\xi_r(y))\right]
 \]
\[
 = t \sum\limits _{u \in \Z ^n}
  ( u_i -b_i) ( u_j - b_j)   P_0(u)
  +
   \sum\limits _{u \in \Z ^n}
  ( u_i -b_i) ( u_j - b_j)  \sum\limits _{r=0}^{t-1}
  \sum\limits _{y \in \Y}
  \P _{\xi}\{ Y _r = y\} c(u,\xi_r(y)).
\]

The statement of the lemma would follow once we show that
for every $s \in \mathbb{S}$
a.s.
\begin{equation}\label{forfeiture}
 \frac{\# \{(r,y): r \leq t, Y _r = y, \xi_r(y) = s \}}{t} = \pi (s).
\end{equation}

Since 
the events $\{ Y _r = y\}$
and $\{ \xi_r(y) = s\}$ are independent,
so by the law of large numbers
\eqref{forfeiture} holds $\P $-a.s.
Hence \eqref{forfeiture} 
also holds
 $\P _\xi$-a.s 
for $\Pi$-a.a. $\xi $, otherwise,
denoting the event of the left hand side of 
\eqref{forfeiture} by $A$, we would have
\[
 \P (A) = \int \P _\xi(A) \pi (d \xi)
 < 1.
\]
 \qed
 
 \textbf{Proof of Theorem \ref{bodacious}}
 Theorem 2.1 by K{\"u}chler and S{\o}rensen
 \cite{KS99} and Lemmas \ref{assuage} and \ref{pull in horns}
 imply that $P _ \xi $ -a.s.
 \begin{equation}
  \frac{1}{t}Y_t \Rightarrow \eta ^2 U,
 \end{equation}
where $U$ a standard $n$-dimensional Gaussian vector.

\bibliographystyle{alpha}
\bibliography{Sinus}

\begin{thebibliography}{BMPZ15}

\bibitem[BC16]{BC16}
G.~Barraquand and I.~Corwin.
\newblock Random-walk in beta-distributed random environment.
\newblock {\em Probability Theory and Related Fields}, pages 1--60, 2016.

\bibitem[Bis11]{surv}
M.~Biskup.
\newblock Recent progress on the random conductance model.
\newblock {\em Probab. Surv.}, 8:294--373, 2011.

\bibitem[BMP94]{BMP94}
C~Boldrighini, R.~A. Minlos, and A.~Pellegrinotti.
\newblock Interacting random walk in a dynamical random environment. {I}.
  {D}ecay of correlations.
\newblock {\em Ann. Inst. H. Poincar\'e Probab. Statist.}, 30(4):519--558,
  1994.

\bibitem[BMP97]{BMP97}
C.~Boldrighini, R.~A. Minlos, and A.~Pellegrinotti.
\newblock Almost-sure central limit theorem for a {M}arkov model of random walk
  in dynamical random environment.
\newblock {\em Probab. Theory Related Fields}, 109(2):245--273, 1997.

\bibitem[BMP07]{BMP07}
K.~Boldrigini, R.~A. Minlos, and A.~Pellegrinotti.
\newblock Random walks in a random (fluctuating) environment.
\newblock {\em Uspekhi Mat. Nauk}, 62(4(376)):27--76, 2007.

\bibitem[BMPZ15]{BMPZ15}
C.~Boldrighini, R.~A. Minlos, A.~Pellegrinotti, and E.~A. Zhizhina.
\newblock Continuous time random walk in dynamic random environment.
\newblock {\em Markov Process. Related Fields}, 21(4):971--1004, 2015.

\bibitem[DP10]{DP10}
L.~Di~Persio.
\newblock Anomalous behaviour of the correction to the central limit theorem
  for a model of random walk in random media.
\newblock {\em Boll. Unione Mat. Ital. (9)}, 3(1):179--206, 2010.

\bibitem[KS99]{KS99}
U.~K{\"u}chler and M.~S{\o}rensen.
\newblock A note on limit theorems for multivariate martingales.
\newblock {\em Bernoulli}, 5(3):483--493, 1999.

\bibitem[TLD17]{TL17}
T.~Thiery and P.~Le~Doussal.
\newblock Exact solution for a random walk in a time-dependent 1d random
  environment: the point-to-point beta polymer.
\newblock {\em Journal of Physics A: Mathematical and Theoretical},
  50(4):045001, 2017.

\bibitem[Zei06]{Zei06}
O.~Zeitouni.
\newblock Random walks in random environments.
\newblock {\em J. Phys. A}, 39(40):R433--R464, 2006.

\end{thebibliography}

\end{document}